\documentclass[letterpaper, 10 pt, conference]{IEEEtran}

\usepackage[noadjust]{cite}
\usepackage{graphicx,xcolor,epstopdf}
\usepackage{amsmath,amsfonts,amssymb,amsthm,mathtools}

\newtheorem{theorem}{Theorem}

\newtheorem{lemma}{Lemma}

\theoremstyle{remark}
\newtheorem*{remark}{Remark}
\newcommand{\R}{\mathbb{R}}

\usepackage{hyperref}
\hypersetup{
	colorlinks,
	linkcolor={violet!50!black},
	citecolor={purple!50!black},
	urlcolor={blue!50!black}
}

\begin{document}

\title{\textit{Robust Stability Analysis of a Class of LTV Systems}}

\author{\IEEEauthorblockN{Shahzad Ahmed\IEEEauthorrefmark{1}, Hafiz Zeeshan Iqbal Khan\IEEEauthorrefmark{1}\IEEEauthorrefmark{2}, and Jamshed Riaz\IEEEauthorrefmark{2}}
\IEEEauthorblockA{\IEEEauthorrefmark{1} Centers of Excellence in Science and Applied Technologies, Islamabad. \\
\IEEEauthorrefmark{2} Department of Aeronautics \& Astronautics, Institute of Space Technology, Islamabad.}}

\maketitle

\begin{abstract}
\textit{\mdseries 
Many physical systems are inherently time-varying in nature. When these systems are linearized around a trajectory, generally, the resulting system is Linear Time-Varying (LTV). LTV systems describe an important class of linear systems and can be thought of as a natural extension of LTI systems. However, it is well known that, unlike LTI systems, the eigenvalues of an LTV system do not determine its stability. In this paper, the stability conditions for a class of LTV systems are derived. This class is composed of piecewise LTV systems, i.e. LTV systems that are piecewise linear in time. Sufficient conditions of stability are derived in the form of linear matrix inequalities (LMIs) by using the Lyapunov stability criterion. The feasibility of LMIs guarantees the stability of a given piecewise LTV system. Furthermore, uncertain piecewise LTV systems with scalar parametric uncertainty are also considered. Sufficient conditions for robust stability of this case are also presented, which come out to be quasi-LMIs, which can be optimized using a bisection algorithm to find the bounds of uncertainty for which the system is stable. The proposed method is applied to the problem of pitch angle control of a space launch vehicle, and the results are presented.
\\}
\end{abstract}


\section{Introduction}
Linear time-varying (LTV) systems emerge from nonlinear time-varying systems and have considerable significance in control design. Many design procedures of adaptive controllers utilize LTV systems as an adaptive closed loop system \cite{gao2018leader,shamma1991guaranteed}. Over the past years, few research works emphasized the stability of LTV systems and some of them discussed various cases of LTI systems \cite{mullhaupt2007numerical,zhou2013stabilisation,mazenc2014reduction,haken1966new}. The evolution of the state space approach and further enhancements in the development of various stability tests for LTI systems were done but at the same time, the development of stability methods for validating the stability of LTV systems was not carried at the same pace due to many reasons \cite{harris1980stability}. One of the main causes can be the complexity of finding the analytical form of state transition matrices of LTV systems. In LTV systems, eigenvalues of the coefficient matrix do not decide the stability. Some linear time-varying systems have eigenvalues of negative real part but they are unstable and vice versa.  However, state transition matrices are required to validate the stability of LTV systems and it is difficult to find the state transition matrices of any time-varying linear system \cite{harris1980stability}. In another case, the stability of LTI systems completely depends upon the position of the poles of the coefficient matrix. LTI system having all of its eigenvalues with negative real parts is termed as a stable system which is not the case with LTV system \cite{rugh1996linear}. Necessary and sufficient conditions of stability for a general class of LTV systems have not been explored much, except for periodic time-varying linear systems \cite{zhou2011periodic,zhou2013and}.

The stability of slowly-varying linear time-varying systems is discussed in \cite{desoer1969slowly,coppel2006dichotomies,rosenbrook1963stability}. The role of constraining the bound on the rate of change of $A(t)$ and the effect of its spectrum on the stability of a system is discussed in detail. Unlike the asymptotic stability of LTI systems, the notion of uniform and non-uniform stability is normally studied for LTV systems. The separation of uniform from non-uniform stability and its linkage with initial time complicates the stability analysis \cite{kalman1960control}. Different stability tools are available in the literature to assess their uniform and non-uniform stability \cite{ramarajan1986time}.

Despite all this, some efforts have been made in the past to find sufficient conditions of stability for time-varying linear plants in some non-conservative fashion. One of the works approximated the time-varying system with a time-invariant linear system and then utilized the Lyapunov stability tool to assess its stability \cite{ilchmann1987sufficient,mullhaupt2007numerical}. Another work treated the time-varying factors as disturbances and took the original time-varying plant as an LTI system and used linear time-invariant system methods to obtain the stability conditions \cite{cao2000computation}. Bellman-Gronwall inequality approach, as given in \cite{bellman2008stability}, is used to obtain the asymptotic stability condition of the LTV system. As stated in \cite{aeyels1999uniform}, if the variation of $A(t)$ lies in between fast and slow modes, then stability analysis of such a system is a challenging task. 

One of the natural extensions of LTI systems could be a class of time-varying linear plants, whose state space matrices are piecewise linear functions of time, which we will refer to as piecewise linear time-varying systems. Such systems naturally arise from linearization of nonlinear time-varying systems, along a trajectory at a given grid of time instants. Those instants are carefully chosen to sufficiently capture the nonlinearities in the original plant. Due to their practical significance, it is essential to analyze the stability properties of piecewise LTV systems. This work focuses on deriving the sufficient conditions of stability of piecewise LTV systems in the form of LMIs. Moreover, robust stability conditions for piecewise LTV closed loop systems in the presence of scalar parametric uncertainty, are obtained. 

The rest of the paper is organized as follows: piecewise LTV system is discussed in the second section along with some stability results for time-varying linear systems, in the third section contributions in the form of main results are given while section four discusses an example of an under-considered system with simulation results. Section five concludes this research work.


\section{Piecewise Linear LTV Systems}
In general, a linear time-varying continuous system can be expressed as:
\begin{equation}
    \dot x = A(t)x(t),\qquad x(t_0) = x_0 \label{eq:1}     
\end{equation}
where $x\in\R^n$ is the state vector, $A(t)$ is the time-varying coefficient matrix, and $x_0$ is the initial state at time $t_0$. Unlike LTI systems, the stability of an LTV system of the form \eqref{eq:1}, can not be determined solely by observing the eigenvalues of the matrix $A(t)$. Like in \cite{wu1974note}, an example of an LTV system having one of its eigenvalues with a positive real part, was proved to be stable. Moreover, the converse case has also been shown true for some other examples of LTV systems \cite{rosenbrook1963stability}. This affirms that the criteria of LTI systems can not be applied to LTV systems for determining their stability. However, the solution in the form of a state transition matrix can be used for evaluating their stability but obtaining the analytical form of state transition matrices for an LTV system is generally very difficult \cite{zhou2016asymptotic}. 

Piecewise LTV systems arise when a nonlinear time-varying system is linearized along a trajectory at a grid of time values. This class of LTV systems can be written as,
\begin{equation}
    \dot x = \mathcal{A}(t)x(t),  \label{eq:2} 
\end{equation}
where $\mathcal{A}(t) = \alpha_k(t) A_k + (1-\alpha_k(t))A_{k-1}$ for all $t \in [t_{k-1},t_k]$ and $\alpha_k(t)$ is given as,
\begin{equation*}
\alpha_k(t) = \dfrac{t-t_{k-1}}{t_k - t_{k-1}},\quad\forall\;k\in[1,N]. 
\end{equation*}
\noindent 
Furthermore, it is assumed that for all $t\leq t_0$, $\mathcal{A}(t) = A_0$, and for all $t\geq t_N$, $\mathcal{A}(t) = A_N$. 

The piecewise behavior in time and emergence from the linearization of real-world nonlinear time-varying systems demand the evaluation of the stability characteristics of such systems. Moreover, the robustness of piecewise LTV systems needs to be assessed in the presence of some parametric uncertainties. In this work, we considered only scalar parametric uncertainty, and the corresponding piecewise LTV can be mathematically expressed as,
\begin{equation}
\dot x = [\mathcal{A}(t)+\delta \mathcal{B}(t)]x(t)  \label{eq:3} 
\end{equation}
where $\delta \in [\frac{1}{\Delta},\Delta]$  is the scalar, uncertain or unknown parameter, where $\Delta \geq 1$ represents its bound. $\mathcal{A}(t)$ and $\mathcal{B}(t)$ in this case are given as:

$$\left.\begin{aligned} \mathcal{A}(t) &= \alpha_k(t) A_k + (1-\alpha_k(t))A_{k-1} \\ 
\mathcal{B}(t) &= \alpha_k(t) B_k + (1-\alpha_k(t))B_{k-1} \end{aligned}\;\right\}\;\;\forall\;t\in[t_{k-1},t_k],$$
\noindent
where $\alpha_k(t) = \dfrac{t-t_{k-1}}{t_k - t_{k-1}}$, $\forall\;k\in[1,N]$. The robust stability of these type of systems is also discussed in next section. Before proceeding further lets briefly recall the S procedure. 

\begin{lemma}[S Procedure]
$z^\top Q z \geq 0$ for all $ z\in S \triangleq \left\{z\in\R^n |  z^\top Gz \geq 0 \right\}$ if there exists a scalar $\tau \geq 0$ 
such that $Q - \tau G \succeq 0$.
\end{lemma}

\section{Main Results}
In this section, the main results of this study are derived in the form of LMI conditions for the stability of piecewise linear time-varying systems. Moreover, robust stability conditions for an uncertain piecewise LTV system are also obtained.
\subsection{Stability Analysis}
For obtaining sufficient conditions of stability for a piecewise LTV system \eqref{eq:2}, the Lyapunov stability criterion is used to derive LMI conditions, which are presented in the following theorem:
\begin{theorem}
A piecewise LTV system \eqref{eq:2} is globally uniformly exponentially stable if the following LMIs are feasible,
\begin{equation}
\begin{split}
\begin{aligned}
    P_0 &\succ 0 \\ P_0 A_0 + A_0^\top P_0 +\varepsilon_0 P_0 &\prec 0 \\ P_N A_N + A_N^\top P_N + \varepsilon_N P_N &\prec 0 
\end{aligned}  \qquad \;\;& \\[1em]  
\left.\begin{aligned}\begin{bmatrix} M_k - \Gamma_k & L_k +\frac{1}{2}\Gamma_k + S_k \\ \star & N_k\end{bmatrix} &\prec 0 \\
\Gamma_k &\succ 0 \\
P_k &\succ 0     \\ 
\varepsilon_k &> 0 \\
S_k^\top &= -S_k\end{aligned}\;\right\}&, \forall\;k\in[1,N]
\end{split}\label{eq:4}
\end{equation}
where,
\begin{equation}
\begin{split} 
M_k &\triangleq \big<P_kA_k - (P_kA_{k-1} + P_{k-1}A_k) + P_{k-1}A_{k-1}\big>_S\\
L_k &\triangleq \bigg<\frac{1}{2}(P_kA_{k-1} + P_{k-1}A_k) - P_{k-1}A_{k-1}\bigg>_S \\ &\quad+ \frac{\varepsilon_k}{2} (P_k-P_{k-1}) \\
N_k &\triangleq \big<P_{k-1}A_{k-1}\big>_S + \gamma_k(P_k - P_{k-1} ) + \varepsilon_k P_{k-1}
\end{split} \label{eq:5a}
\end{equation}
where $\left<\cdot\right>_S \triangleq (\cdot) + (\cdot)^\top$, and $\gamma_k \triangleq \dot{\alpha}_k = (t_k - t_{k-1})^{-1}$.

Here $\Gamma_k$ and $P_k$ are positive definite matrices, $S_k$ is skew symmetric matrix, and $\varepsilon_k > 0$ represents some prescribed values which corresponds to the exponential decay rates. 

\end{theorem}
\begin{proof}
 Let's consider the following Lyapunov candidate function:
\begin{equation}
\begin{split}
    \mathcal{V} = x^\top \mathcal{P}(t) x    
\end{split} \label{eq:6}
\end{equation}
where 
\begin{equation}
\mathcal{P}(t) = \alpha_k(t) P_k + (1-\alpha_k(t))P_{k-1} \label{eq:7}
\end{equation}
for all $t\in[t_{k-1},t_k]$, and  for all $k$. For $\mathcal{V} >0$, we need $P_k \succ 0$ for all $k\in[0,N]$.
Consider the time interval $t\in[t_{k-1},t_k]$, then
\begin{equation}
\begin{split}
\dot{\mathcal{V}} &= x^\top [\mathcal{P}(t)\mathcal{A}(t)+\mathcal{A}(t)^\top \mathcal{P}(t) +\dot{\mathcal{P}}(t)]x\\
&= x^\top \Bigg[\bigg<\alpha_k^2 P_k A_k + \alpha_k(1-\alpha_k) (P_k A_{k-1} + P_{k-1}A_k) \\
&{}+ (1-\alpha_k)^2  P_{k-1}A_{k-1}\bigg>_S +  \gamma_k (P_k - P_{k-1})\Bigg]x 
\end{split} \label{eq:8}
\end{equation}  
So for $\dot{\mathcal{V}} < -\varepsilon_k \mathcal{V}$,  lets denote
\begin{equation}
\dot{\mathcal{V}} + \varepsilon_k \mathcal{V} \triangleq x^\top Q_k x < 0
\label{eq:9} 
\end{equation}
For all $\alpha_k \in [0,1]$, which can be written as:
\begin{equation}    
Q_k =\alpha_k^2 M_k + 2\alpha_k L_k + N_k \prec 0    
\label{eq:10}
\end{equation}
where $M_k$, $L_k$, and $N_k$ are defined in Eq. \eqref{eq:5a}. Let $z=\begin{bmatrix}\alpha_k I & I\end{bmatrix}^\top$, then we can write $Q_k = z^\top R_k z$, where
$$R_k = \begin{bmatrix}M_k & L_k \\ \star & N_k\end{bmatrix}$$
Let $\Gamma_k \succ 0$, then $z^\top G_k z = \alpha_k(1-\alpha_k)\Gamma_k \succ 0$, where \[G_k = \begin{bmatrix}-\Gamma_k & \frac{1}{2} \Gamma_k \\ \star & 0\end{bmatrix}\]
Then using S procedure, $Q_k \prec 0$, if $R_k + \tau_k G_k \prec 0$ for any $\tau_k >0$, since $G_k$ is free, so with any loss of generality, we can choose $\tau_k =1$. This results in the LMI conditions \eqref{eq:4}. Here skew-symmetric matrix $S_k$ is added to cater for the structure of $z$.
\end{proof}
\subsection{Robust Stability Analysis}
For evaluating the robust stability of uncertain piecewise LTV system \eqref{eq:3} in the presence of bounded scalar parametric uncertainty, sufficient conditions are derived in the form of LMIs as described in the following theorem:
\begin{theorem}
The uncertain piecewise LTV system \eqref{eq:3} is globally uniformly exponentially stable for all $\delta\in[1/\Delta, \Delta]$, if following LMIs are feasible,
\begin{equation}
\begin{split}
\begin{aligned}P_0 &\succ 0 \\ P_0 A_0 + A_0^\top P_0 +\varepsilon_0 P_0 &\prec 0 \\ P_N A_N + A_N^\top P_N + \varepsilon_N P_N &\prec 0 \end{aligned}\qquad\qquad& \\
\left.\begin{aligned}R_k + H_k + S_{2,k} &\prec 0 \\
P_k &\succ 0 \\
\Gamma_k &\succ 0 \\
\Psi_k &\succ 0 \\
\varepsilon_k &> 0 \\
S_{1,k}^\top &= -S_{1,k} \\
S_{2,k}^\top &= -S_{2,k}\end{aligned}\;\right\}&, \forall\;k\in[1,N]
\end{split}\label{eq:11}
\end{equation}
where
\begin{equation}
\begin{split}
R_k &\triangleq \begin{bmatrix}0 & \frac{1}{2}N_{2,k} & 0 & \frac{1}{4}N_{1,k} \\ \star & M_{2,k} & \frac{1}{4}N_{1,k}^\top & \frac{1}{2} M_{1,k} \\ \star & \star & -\Gamma_k & \frac{1}{2}N_{0,k} + \beta \; \Gamma_k + S_{1,k}\\ \star & \star & \star & M_{0,k}-\Gamma_k\end{bmatrix}    \\
H_k &\triangleq \begin{bmatrix}-\Psi_k & \frac{1}{2} \Psi_k  \\ \star & 0 \end{bmatrix} 
\end{split}\label{eq:12}
\end{equation}
where
\begin{equation}
\begin{split} 
M_{2,k} &\triangleq \big<P_kA_k - (P_kA_{k-1} + P_{k-1}A_k) + P_{k-1}A_{k-1}\big>_S \\
M_{1,k} &\triangleq \bigg<\frac{1}{2}(P_kA_{k-1} + P_{k-1}A_k) - P_{k-1}A_{k-1}\bigg>_S \\
&{}\qquad + \frac{\varepsilon_k}{2} (P_k-P_{k-1}) \\
M_{0,k} &\triangleq \big<P_{k-1}A_{k-1}\big>_S + \gamma_k(P_k - P_{k-1} ) + \varepsilon_k P_{k-1} \\
N_{2,k} &\triangleq \big<P_kB_k - (P_kB_{k-1} + P_{k-1}B_k) + P_{k-1}B_{k-1}\big>_S \\
N_{1,k} &\triangleq \bigg<\frac{1}{2}(P_kB_{k-1} + P_{k-1}B_k) - P_{k-1}B_{k-1}\bigg>_S \\
N_{0,k} &\triangleq \big<P_{k-1}B_{k-1}\big>_S \end{split}\qquad\raisetag{5\baselineskip}\label{eq:13}
\end{equation}
Here $\Gamma_k$, $\Psi_k$ and $P_k$ are positive definite matrices, $S_{1,k}$ are $S_{2,k}$ are skew symmetric matrices, $\beta = \frac{1}{2}\left(\Delta + \frac{1}{\Delta}\right)$, and $\varepsilon_k > 0$ represents some prescribed values which corresponds to the exponential decay rates. 
\end{theorem}

\begin{proof}
Consider the same Lyapunov function Eq. \eqref{eq:6}. For $\mathcal{V} >0$, we need $P_k \succ 0$ for all $k\in[0,N]$.
Consider the time interval $t\in[t_{k-1},t_k]$, then we get
\begin{equation}
\begin{split}\dot{\mathcal{V}} &= x^\top [\left<\mathcal{P}(t)\mathcal{A}(t) + \delta \mathcal{P}(t)\mathcal{B}(t)\right>_S  +  \dot{\mathcal{P}}(t)]x \\ 
&=x^\top \Bigg[\bigg<\alpha_k^2 P_k A_k + \alpha_k(1-\alpha_k) (P_k A_{k-1} + P_{k-1}A_k) \\
&{}\quad+ (1-\alpha_k)^2 P_{k-1}A_{k-1}\bigg>_S  
+\delta\;\bigg<\alpha_k^2 P_k B_k \\
&{}\qquad+ \alpha_k(1-\alpha_k)(P_k B_{k-1} + P_{k-1}B_k) \\
&{}\qquad+ (1-\alpha_k)^2 P_{k-1}B_{k-1}\bigg>_S +  \gamma_k (P_k - P_{k-1})\Bigg]x
\end{split} \label{eq:15}\raisetag{4\baselineskip}
\end{equation}
So for $\dot{\mathcal{V}} < -\varepsilon_k \mathcal{V}$,  lets denote
\begin{equation}
\dot{\mathcal{V}} + \varepsilon_k \mathcal{V} \triangleq x^\top \hat{Q}_k x < 0
\label{eq:16}
\end{equation}
For all $\alpha_k \in [0,1]$ and $\delta \in [\frac{1}{\Delta},\Delta]$, which can be written as:
\begin{equation}
\begin{split}
\hat{Q}_k &= \mathcal{M}_k(\alpha_k) + \delta\;\mathcal{N}_k(\alpha_k) \\
&\triangleq \bigg(\alpha_k^2 M_{2,k} + 2\alpha_k M_{1,k} + M_{0,k}\bigg) \\ 
&{}\qquad+ \delta \bigg(\alpha_k^2 N_{2,k} + 2\alpha_k N_{1,k} + N_{0,k}\bigg) \prec 0
\end{split} \label{eq:17}
\end{equation}
where $M_{i,k}$ and $N_{i,k}$ for $i\in[0,2]$ are defined in Eqs. \eqref{eq:13}. Let $z=\begin{bmatrix}\delta I & I\end{bmatrix}^\top$ , then $\hat{Q}_k = z^\top \hat{R}_k(\alpha_k) z$, where

\begin{equation}
\hat{R}_k(\alpha_k) = \begin{bmatrix}0 & \frac{1}{2}\mathcal{N}_k(\alpha_k) \\ \star & \mathcal{M}_k(\alpha_k)\end{bmatrix}  
\end{equation}

Then using S procedure, $\hat{Q}_k \prec 0$ for all $\delta \in [\frac{1}{\Delta},\Delta]$, if $\hat{Q}_k + \tau_k g_k < 0$ for some $\tau_k> 0$. where $g_k = (\delta - \frac{1}{\Delta})(\Delta-\delta) \geq0$. Let $\Gamma_k \succ 0$, So then $\tau_k g_k = z^\top G_k z = (\delta - \frac{1}{\Delta})(\Delta-\delta)\Gamma_k \succ 0$, where 
\begin{equation}
G_k = \begin{bmatrix}-\Gamma_k & \beta\;\Gamma_k \\ \star & -\Gamma_k\end{bmatrix}    
\end{equation}
where $\beta = \frac{1}{2}\left(\Delta + \frac{1}{\Delta}\right)$, so we need $Q_k(\alpha_k) \triangleq \hat{R}_k(\alpha_k) + G_k \prec 0$, which gives, 
\begin{equation}
Q_k(\alpha_k) =\begin{bmatrix}-\Gamma_k & \frac{1}{2}\mathcal{N}_k(\alpha_k)+\beta\;\Gamma_k + S_{1,k} \\ \star & \mathcal{M}_k(\alpha_k)-\Gamma_k\end{bmatrix} \prec 0  
\end{equation}
Let $\hat{z}=\begin{bmatrix}\alpha_kI & I\end{bmatrix}^\top$ , then $Q_k = \hat{z}^\top R_k \hat{z}$, where $R_k$ is defined in Eq. \eqref{eq:12}. Then using S procedure again, $Q_k(\alpha_k) \prec 0$ for all $\alpha_k \in [0,1]$, if $Q_k + \tau_k g_k < 0$ for some $\tau_k > 0$, where $g_k = \alpha_k(1-\alpha_k)$. Let $\Psi_k \succ 0$. So then $\tau_k g_k = \hat{z}^\top H_k \hat{z} = \alpha_k(1-\alpha_k)\Psi_k \succ 0$, where $H_k$ is defined in Eq. \eqref{eq:12}. This results in LMI conditions \eqref{eq:11} for robust stability. The skew-symmetric matrices $S_{1,k}$ and $S_{2,k}$ are added to cater for the structure of $z$ and $\hat{z}$, respectively.
\end{proof}

\begin{remark}
It must be noted that for a given $\Delta$ (or $\beta$) Eq. \eqref{eq:11} is an LMI condition which can be easily checked for feasibility. 
\end{remark}

\begin{remark}
Since $\beta = \frac{1}{2}\left(\Delta + \frac{1}{\Delta}\right)$ is a strictly increasing function of $\Delta$ for all $\Delta > 1$. Therefore, to find the maximum uncertainty region ($[1/\Delta,\Delta]$), we can maximize $\beta$ subject to Eq. \eqref{eq:11}. Which result in a quasi-LMI condition, and can be easily solved using bisection.
\end{remark}

\section{An Illustrative Example}

\begin{figure}
    \centering
    \includegraphics[width=\linewidth]{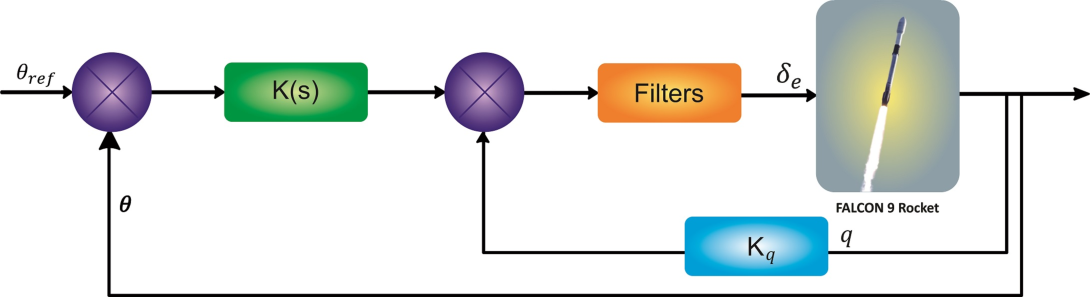}
    \caption{Control architecture}
    \label{Figure:1}
\end{figure}

\begin{figure}
    \centering
   \includegraphics[width=\linewidth]{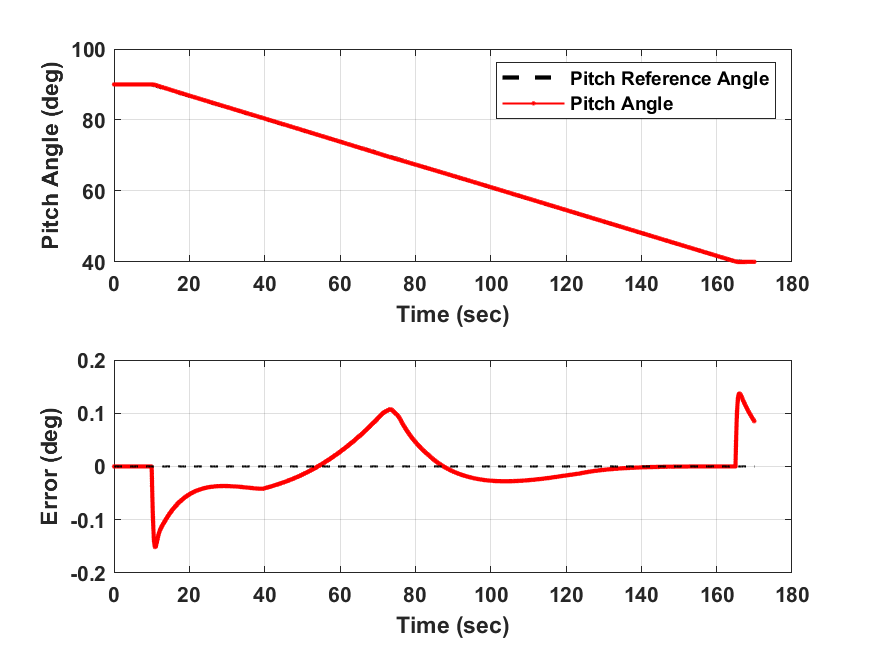}
    \caption{Response of pitch angle}
    \label{Figure:2}
\end{figure}

The proposed approach is applied to pitch control of \emph{Falcon 9} space launch vehicle (SLV). As given in \cite{aamer2023modelling}, linear time-varying models over the trajectory until stage separation, are selected and their control architecture is considered. The block diagram of the closed-loop system is shown in Figure \ref{Figure:1}. Pitch angle and rates are the output of the under-considered SLV system. $\theta_{ref}$ is the pitch reference while $\delta_e$ represents the elevator deflections. The selected models account for flexible body dynamics over the selected flight envelope. During modeling, the earth is assumed to be non-rotating and flat. ``Tails wags dog'' and sloshing effects are ignored to reduce complexity. In the nominal scenario, the stability criteria given in   Eq. \eqref{eq:4} is found to be feasible using YALMIP and SeDuMi toolboxes \cite{lofberg2004yalmip,sturm1999using}, thus the system is stable. The reference tracking response of pitch angle and its error is shown in Figure \ref{Figure:2}. It can be seen that the pitch angle of the vehicle tracks the reference trajectory with negligible error, and the stability is preserved throughout the flight. 

\begin{figure}
    \centering
   \includegraphics[width=0.9\linewidth]{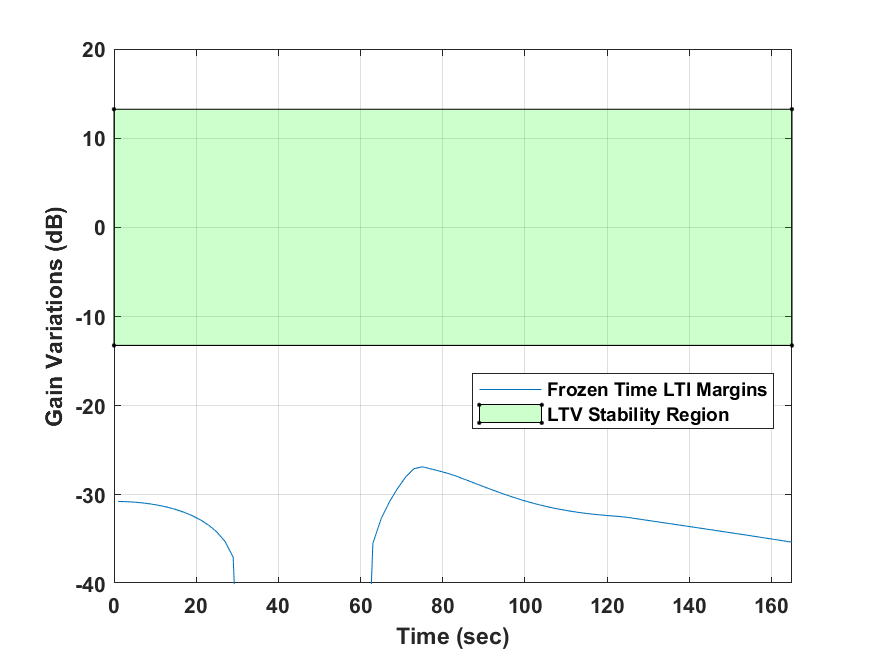}
    \caption{Frozen-time LTI vs. proposed LTV margins}
    \label{Figure:3}
\end{figure}

\begin{figure}
    \centering
   \includegraphics[width=\linewidth]{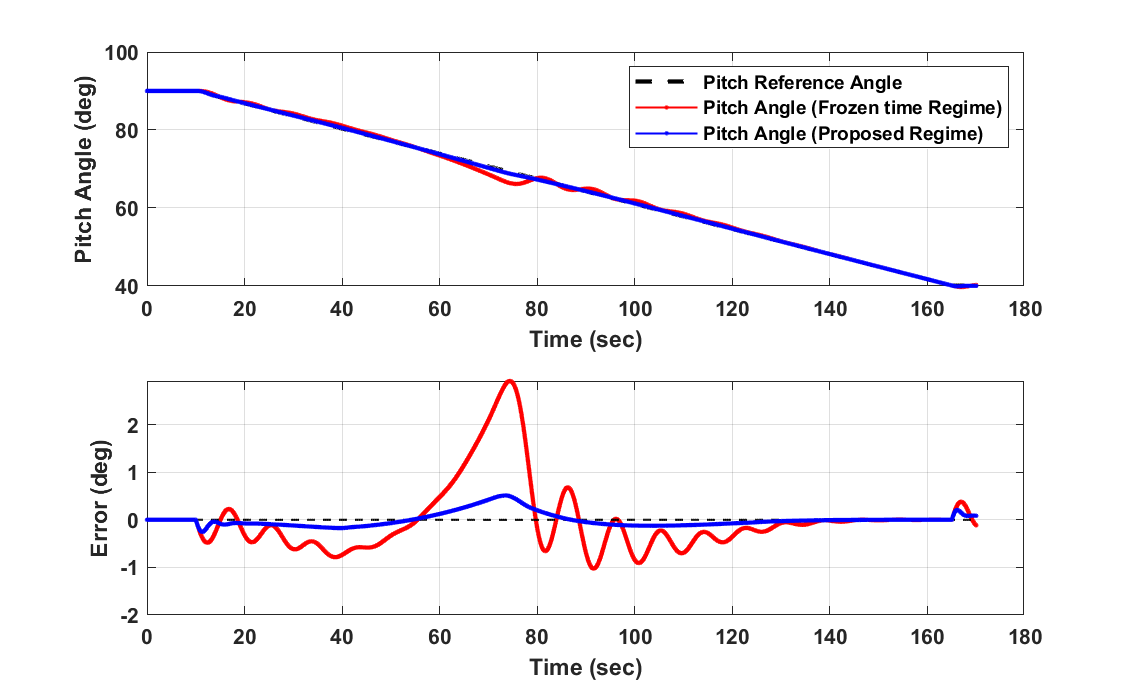}
    \caption{Tracking performance comparison}
    \label{Figure:4}
\end{figure}

Then robust stability conditions are tested to find the maximum region of the uncertainty at the input of the system or equivalently the gain margin. The quasi-LMIs in \eqref{eq:11} are solved by using a bisection method to maximize $\beta$. This yields a lower margin of $0.2176$, and an upper margin of $4.5956$. More precisely, it concludes that the closed-loop system is stable for all $\delta \in [0.2176,4.5956]$. To prove its effectiveness, the proposed scheme is compared with the frozen-time-based LTI gain margins. The frozen time approach gives a maximum lower gain margin of $0.05$. The variation of lower gain margin given by the frozen time approach is shown in Figure \ref{Figure:3}, along with the stability region (green shaded region) predicted by the proposed approach. Moreover, as shown in Figure \ref{Figure:4}, it can be seen that if the gain at the input is perturbed up to the lower margin provided by the proposed robust stability analysis, the closed-loop response remains stable and shows good performance as well. However, if it is perturbed up to the value computed using the frozen-time approach, the resulting closed-loop response has significantly large oscillations and poor performance. Overall results emphasized the fact the frozen-time approaches are not reliable in the case of LTV systems. However, the proposed method not only provides stability and robustness guarantees but also applicable to a wide class of LTV systems. 

\section{Conclusion}
Keeping in view the complexity of conventional stability methods of LTV systems, in this work stability analysis of a class of LTV systems is considered. Sufficient conditions of stability are formulated in the form of LMIs. It is proved that the feasibility of derived LMIs guarantees the stability of a given piecewise LTV system. The conditions for robust stability in the presence of bounded scalar parametric uncertainty are also derived. The proposed framework makes use of the bisection method to find the maximal interval of uncertainty for which the perturbed system is stable. To demonstrate the applicability of the proposed approach an example of pitch control of an SLV is considered and its results are presented and compared with the frozen-time LTI method.

\bibliography{References}


\end{document}